\documentclass{article}
\usepackage{spconf,amsmath,graphicx}
\usepackage{amsfonts, amssymb}
\usepackage{url}
\usepackage{cite}
\usepackage{hyperref}
\usepackage{array,eqparbox,collcell}
\usepackage{gensymb}

\usepackage{xcolor}

\def\z{{\mathbf z}}

\title{HRTF Field: Unifying Measured HRTF Magnitude Representation with Neural Fields}
\name{You Zhang, Yuxiang Wang, Zhiyao Duan \thanks{This work is partially supported by the Goergen Institute for Data Science at the University of Rochester, a New York State Center of Excellence in Data Science award, and synergistic activities funded by the National Science Foundation (NSF) under grant DGE-1922591. The authors would also like to thank Prof. Mark Bocko for brief discussions.
}}
\address{University of Rochester, Rochester, NY, USA}

\begin{document}
\ninept
\maketitle
\begin{abstract}
Head-related transfer functions (HRTFs) are a set of functions describing the spatial filtering effect of the outer ear (i.e., torso, head, and pinnae) onto sound sources at different azimuth and elevation angles. They are widely used in spatial audio rendering. While the azimuth and elevation angles are intrinsically continuous, measured HRTFs in existing datasets employ different spatial sampling schemes, making it difficult to model HRTFs across datasets. In this work, we propose to use neural fields, a differentiable representation of functions through neural networks, to model HRTFs with arbitrary spatial sampling schemes. Such representation 
is unified across datasets with different spatial sampling schemes. HRTFs for arbitrary azimuth and elevation angles can be derived from this representation. We further introduce a generative model named \textit{HRTF field} to learn the latent space of the HRTF neural fields across subjects. We demonstrate promising performance on HRTF interpolation and generation tasks and point out potential future work.

\end{abstract}
\section{Introduction}
\label{sec:intro}
The geometry of a listener's auditory apertures (i.e., pinnae, head and upper torso) imposes spatial filtering effects onto the sound propagating from the sound source to the ear canal. Such effects are mathematically characterized by head-related transfer functions (HRTFs), which are widely used in headphone-based spatial audio rendering~\cite{xie2013head}. Existing spatial audio rendering techniques mostly use a generic HRTF (measured on a dummy head) and often lead to sound localization confusion~\cite{armstrong2018perceptual}. This is because the geometry
differs significantly from one person to another, and the corresponding HRTFs also differ. Modeling HRTFs of different listeners is a critical step to predicting HRTFs from the geometry of a new listener, for personalized spatial audio rendering. This can significantly improve the immersiveness in virtual and augmented reality. 

The individual HRTF measurement process requires an anechoic room with microphones inserted in the listeners' ears to record the head-related impulse responses (HRIRs). Typically, multiple loudspeakers are arranged around an arc and then rotating in another orthogonal direction~\cite{app10145014, algazi2001cipic}. Such a time-consuming and cumbersome process makes it impractical to measure each user's HRTF. Thus, it is necessary to predict one's HRTF based on geometry. Some methods have been developed for numerically calculated HRTF~\cite{pollack2022modern}, but they suffer from unrealistic assumptions. Recent efforts have been put into data-driven methods, using machine learning to perform HRTF personalization~\cite{cobos2022overview} from anthropometric measurements~\cite{grijalva2016manifold, chen2019autoencoding, wang2021global}, ear images~\cite{lee2018personalized, zhao2022magnitude}, or scanned head mesh~\cite{miccini2020hrtf, wang2022predicting}.



\begin{figure}[t]
\begin{minipage}[htbp]{1.0\linewidth}
  \centering
  \centerline{\includegraphics[width=1.01\columnwidth]{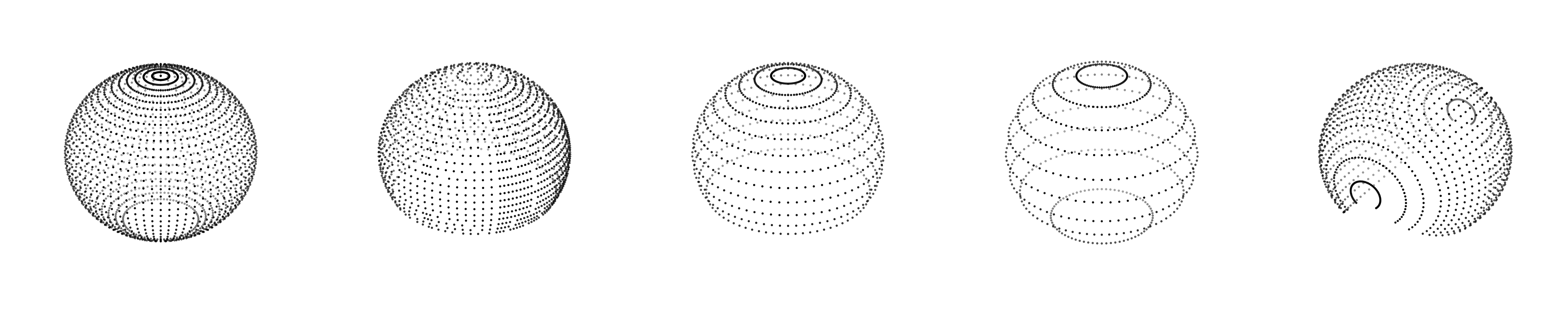}}
\vspace{-0.5pt}
  \small
  \centerline{\hspace{0pt}Aachen\hspace{28pt} ARI\hspace{34pt} RIEC\hspace{30.5pt} 3D3A\hspace{27.9pt} CIPIC}\medskip
\vspace{-0.5pt}
\end{minipage}

\begin{minipage}[htbp]{1.0\linewidth}
  \centering
  \centerline{\includegraphics[width=1.01\columnwidth]{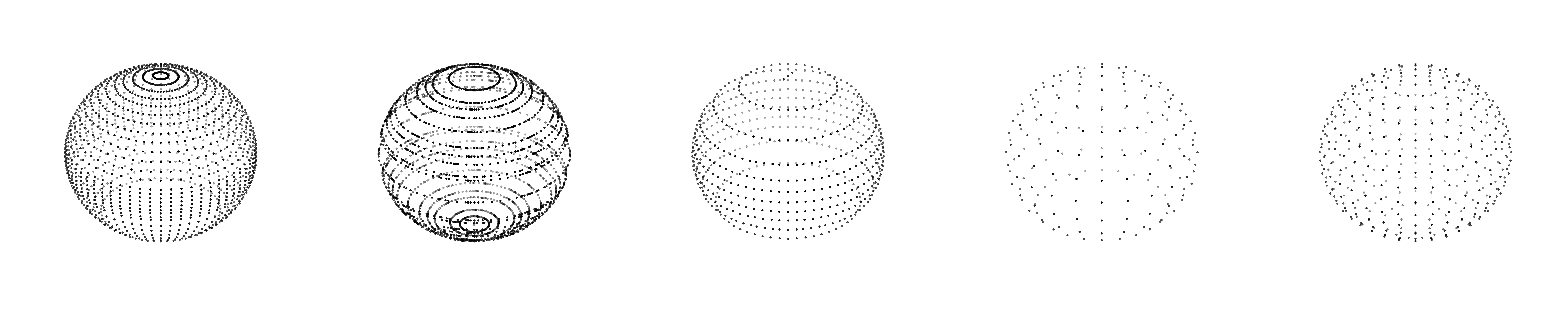}}
\vspace{-0.5pt}
  \small
  \centerline{\hspace{13pt}BiLi\hspace{24.5pt} SADIE II\hspace{15.5pt} Crossmod\hspace{21pt} Listen\hspace{22pt} HUTUBS}\medskip
\end{minipage}
\caption{Configurations of sound source positions used in existing publicly-available measured far-field HRTF databases.}
\label{fig:sample}
\end{figure}

\begin{figure}[]
  \centering
  \centerline{\includegraphics[width=0.97\columnwidth]{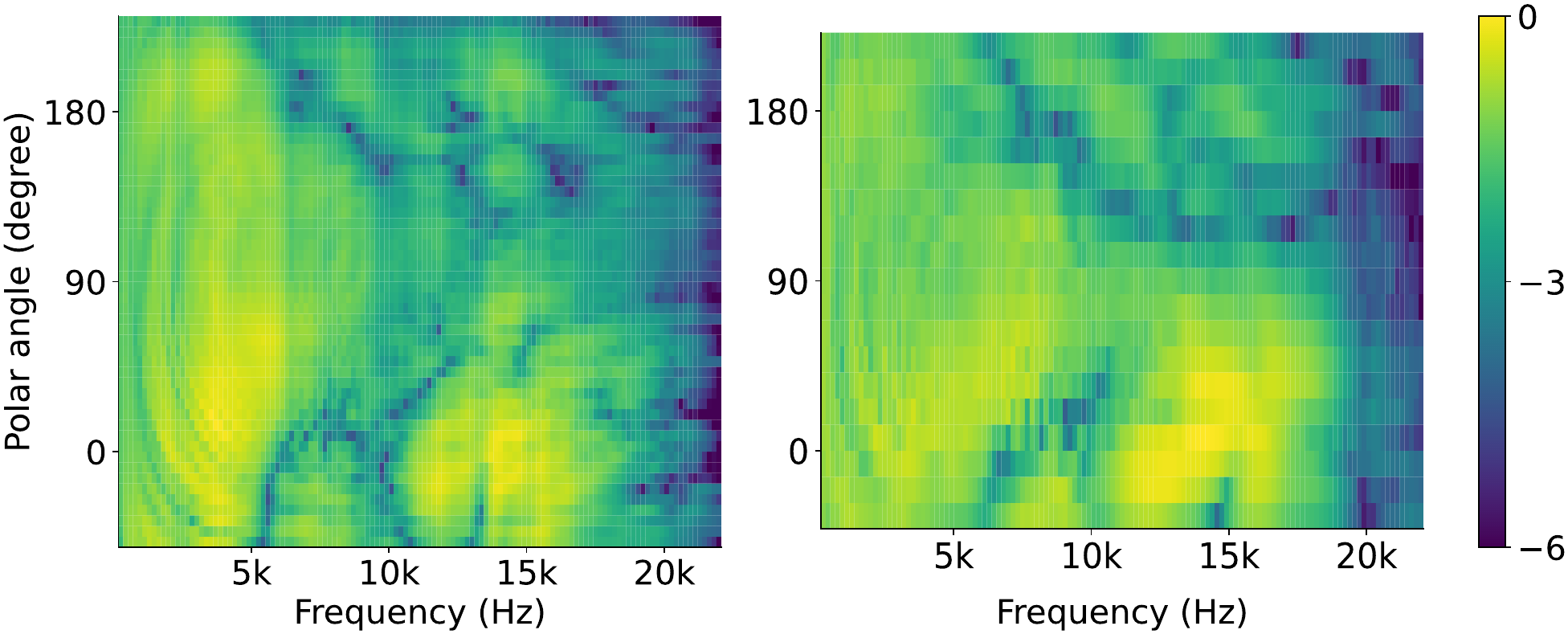}}
\vspace{-0.5pt}
\caption{The left ear HRTF magnitudes (dB) of the midsagittal planes of two subjects sampled from the CIPIC (left) and the Listen (right) dataset, respectively. The color represents the magnitude in the log scale, where 0 dB corresponds to the maximum magnitude.}
\label{fig:midsagittal}
\end{figure}

However, these methods have all been trained on small data (dozens of subjects) with overfitting risks, as their data representation is locked to a specific spatial sampling scheme of the HRTF training data and cannot utilize data with different sampling schemes across datasets. As shown in Fig.~\ref{fig:sample}, the source location grid used in existing publicly available measured HRTF datasets differs from one to another, which also restricts the format of predicted HRTFs to the given source grids. The spherical coordinate systems are different. 
While the left four columns of datasets use geodesic representations, CIPIC uses the interaural-polar coordinate system, and HUTUBS uses the near-Lebedev system. 
Different sampling schemes and coordinate systems cause discrepancy in spatial resolution of the HRTF measurements. For example, the midsagittal HRTF pattern is compared between the CIPIC and the Listen dataset in Fig.~\ref{fig:midsagittal}. While the overall pattern is similar, as they should be, the different spatial resolutions make it difficult to compare the differences in peaks and notches between the two listeners, which are critical and personalized localization cues for spatial audio rendering~\cite{asano1990role}. This motivates us to seek a unified representation of HRTFs across different datasets to alleviate the bounds of specific spatial sampling schemes.

One grid-agnostic solution to represent the HRTFs on a spherical surface is using spherical harmonics basis functions, as explored in~\cite{evans1998analyzing, zotkin2009regularized, 7096941}. However, it requires a near-uniform sampling scheme and a complete coverage of the spherical surface, such as in the HUTUBS database~\cite{fabian2019hutubs}; otherwise, there will be aliasing~\cite{zotkin2009regularized}. Most of the existing measured databases, as shown in Fig.~\ref{fig:sample}, lack the sampled source positions at the bottom of the sphere due to the inconvenience of measurements and the futility of HRTFs at such locations. Another work~\cite{gebru2021implicit} investigated modeling personalized HRTFs implicitly by estimating the transformation functions for binaural synthesis with neural networks. Their method is also not constrained by the measurement directions, but they directly predict binaural audio where HRTFs are intermediate outputs with no ground truth.

In this work, we propose a magnitude modeling approach for measured HRTFs with different spatial sampling schemes using neural fields~\cite{xie2022neural}. We aim to learn a unified and differentiable representation of HRTFs with respect to azimuth and elevation angles.  We further introduce \textit{HRTF field}, a generative model to learn a latent space of the neural fields and apply it to interpolation and generation tasks. Our unified representation with neural fields, for the first time, enables training across existing datasets. We demonstrate that the HRTF field has advantages for interpolation in comparison with existing mathematical methods and is able to generate reasonable synthesized HRTFs. Our work is reproducible with code at \url{https://github.com/yzyouzhang/HRTF_field}.

\section{HRTF Field}

\subsection{Representing HRTFs of a single subject with a neural field}

HRTFs of one ear of a subject is a functional defined on the unit sphere around the subject. It maps the sound-source direction (i.e., azimuth and elevation) to a function of frequency (i.e., frequency responses)~\cite{moller1995head}. When only the magnitude of frequency responses are considered, this mapping can be represented as:
\begin{equation}
\operatorname{HRTF}(\theta, \phi)=\frac{\mathbf{p}\left(\theta, \phi\right)}{\mathbf{p}_0},
\label{eq:hrtf}
\end{equation}
$\mathbf{p}(\theta, \phi)$ is the magnitude spectrum of the sound received at the entrance of ear canal when the sound source is placed at the direction $(\theta, \phi)$. $\mathbf{p}_0$ is the magnitude spectrum of the source signal.


As the HRTF functional is defined on the continuous sphere, we propose to represent it with a neural field~\cite{xie2022neural}, which employs deep neural networks to parameterize a function defined in a continuous domain. Unlike the discrete locations of measured HRTFs, neural fields are agnostic to the grid resolution, and are automatically differentiable with respect to the spatial coordinates. They are also referred to as implicit neural representations~\cite{sitzmann2020implicit} in the literature as the original discrete representation of the magnitudes can be queried by coordinates as input. With memory efficiency and high fidelity, neural fields have been applied to various signal and scene representations, such as images~\cite{tancik2020fourier, sitzmann2020implicit}, shapes~\cite{chen2019learning}, textures~\cite{Oechsle_2019_ICCV}, impulse responses for acoustics propagation~\cite{luo2022learning}, and scene reconstruction for view synthesis~\cite{mildenhall2021nerf}.

For HRTFs of each subject $\operatorname{HRTF}(\theta, \phi)$, we can use a neural field $F(\theta, \phi)$ to model it. The transformation of the network is $F: \mathbb{R}^{2} \mapsto \mathbb{R}^{K}$, where $K$ is the number of frequency bins.
We adopt the sinusoidal representation network architecture (SIREN)~\cite{sitzmann2020implicit}, which is a multi-layer perceptron (MLP) with sine activation functions. Leveraging periodic activation functions, it achieves impressive results with the advantage of representing complicated signals in various domains.
The SIREN network can be described as
\begin{equation}
\begin{aligned}
F(\theta, \phi)=\mathbf{W}_n & \left(g_{n-1} \circ g_{n-2} \circ \ldots \circ g_0(\theta, \phi)\right)+\mathbf{b}_n, \\
\text{where} \quad
 & g_i\left(\mathbf{h}_i\right)= \sin \left(\mathbf{W}_i \mathbf{h}_i+\mathbf{b}_i\right).
\end{aligned}
\label{eq:siren}
\end{equation}
The transform $g_i$ for the $i$-th layer takes the hidden feature $\mathbf{h}_i$ as input and applies a linear transform with weight matrix $\mathbf{W}_i$ and bias $\mathbf{b}_i$, followed by a sine activation except the last layer.


\subsection{Learning HRTF representations across subjects} 

Eq.~\eqref{eq:hrtf} models the HRTFs of a single subject with a neural field. To model HRTFs across multiple subjects, we introduce a $D$-dimensional latent vector $\mathbf{z}$ to the input of the neural field. This vector is meant to correspond to different subjects, and can serve as the connection between a subject's head geometry and their HRTFs.



With this modification, we define \textit{HRTF field} as a mapping from the sound direction $(\theta, \phi)$ and the latent code $\mathbf{z}$ to an individual's HRTFs $\operatorname{HRTF}(\theta, \phi)$. 
A straightforward solution is to use a hypernetwork to generate the network weights of the neural field as in~\cite{sitzmann2020implicit}. 
However, this requires another component and has many hyperparameters to adjust. 
We propose to use the implicit gradient origin network (IGON) proposed in~\cite{bond2020gradient} as the generative model for neural fields. It can be achieved with a simpler overall architecture.
IGON uses the SIREN network architecture, but it takes an additional parameter $\mathbf{z}$ as input, thus can be denoted as $G(\theta, \phi, \mathbf{z})$. The output is HRTF magnitudes across all frequencies at that location for the subject. The mapping becomes $G: \mathbb{R}^{2+D} \mapsto \mathbb{R}^{K}$.

The training procedure of IGON involves co-optimizing the latent space and the network weights. The latent code $\mathbf{z}$ is initialized at the origin $\mathbf{z}_0$ of the latent space. The IGON algorithm first infers $\mathbf{z}$ by computing the gradient of the mean square error (MSE) between the generated and the ground-truth HRTFs of the subject at all directions $\mathbf{x}$ with respect to the origin $\mathbf{z}_0$,
\begin{equation}
\mathbf{z} = \mathbf{z}_0 - \nabla_{\mathbf{z}_0} \mathcal{L}_{\mathrm{MSE}}\left(\mathbf{x}, G\left(~\boldsymbol{\cdot}, ~\boldsymbol{\cdot}, \mathbf{z}_0\right)\right).
\label{eq:z0}
\end{equation}
The next step is to update $G$ by fitting the HRTFs with the MSE loss,
\begin{equation}
\mathcal{L}=\mathcal{L}_{\mathrm{MSE}}\left(\mathbf{x}, G\left(~\boldsymbol{\cdot}, ~\boldsymbol{\cdot}, \mathbf{z}\right)\right).
\label{eq:update}
\end{equation}
Then the training scheme iterates between Eq.~\eqref{eq:z0} and Eq.~\eqref{eq:update} with mini-batches until convergence.
Initializing $\mathbf{z}$ at the origin $\mathbf{z}_0$ is based on the assumption that the mean of the latent distribution is $\mathbf{0}$. This decreases the input variation and thus simplifies the optimization procedure.
As the overall HRTFs have similar pattern for different subjects, we assume HRTFs across subjects lie on a manifold.
We apply IGON to capture the manifold information over the neural fields of the measured HRTF magnitudes across subjects. 

\section{Experimental Setup}
\subsection{Datasets}

The publicly-available datasets of measured far-field HRTFs for human subjects are all available in the spatially oriented format for acoustics (SOFA) repository\footnote{\href{https://www.sofaconventions.org/mediawiki/index.php/Files}{\texttt{https://www.sofaconventions.org}}}. Details can be found in Table~\ref{tab:dataset}. Some particular subjects in each of the datasets may not have their complete set of locations measured, resulting in less number of locations than that shown in the table, then we just use all the locations they provided. Each dataset is measured with a different setup, resulting in different spherical sampling schemes, as shown in Fig.~\ref{fig:sample}. 

\begin{table}[]
\def\splitrange$#1,#2${%
  \eqmakebox[tabcol-l-\thetable][l]{$#1$}%
  $,$%
  \eqmakebox[tabcol-r-\thetable][l]{$#2$}%
}
\newcommand{\setcolentryB}[1]{%
  \expandafter\splitrange#1%
}
\renewcommand{\arraystretch}{1.0}
\centering
\caption{Statistics of public measured far-field HRTF databases. \quad \\
\newline}
\begin{tabular}{l|rr >{\collectcell\setcolentryB}c<{\endcollectcell}}
\hline \hline
Name   & \# Subjects & \# Locations & \multicolumn{1}{l}{Elevation Range} \\ 
\hline
3D3A~\cite{sridhar2017database}   & $38$~~~~~          & $648$~~~~~      &  $[-57\degree\ , \ 75\degree~~~]$  \\
Aachen~\cite{bomhardt2016high} & $48$~~~~~          & $2304$~~~~~    &   $[-66.24\degree\ , \ 90\degree~~~]$  \\
ARI     & $97$~~~~~         & $1550$~~~~~    &   $  [-30\degree\ , \ 80\degree~~~ ] $\\  
BiLi~\cite{carpentier2014measurement}   & $52   $~~~~~       & $1680$~~~~~    &   $[-50.5\degree\ , \ 85.5\degree]$  \\
CIPIC~\cite{algazi2001cipic}  & $45  $~~~~~        & $1250$~~~~~    &    $[-50.62\degree\ , \ 90\degree~~~]$\\ 
Crossmod & $ 24$~~~~~          & $651$~~~~~    &    $[-40\degree\ , \ 90\degree~~~]$  \\  
HUTUBS~\cite{fabian2019hutubs}   & $96$~~~~~          & $440$~~~~~    &     $[-90\degree\ , \ 90\degree~~~]$ \\
Listen   & $50$~~~~~          & $187$~~~~~    &      $[-45\degree\ , \ 90\degree~~~]$\\   
RIEC~\cite{watanabe2014dataset}     & $105$~~~~~          & $865$~~~~~    &     $[-30\degree\ , \ 90\degree~~~]$ \\
SADIE II~\cite{armstrong2018perceptual}    & $18$~~~~~          & $2818$~~~~~    &   $[-90\degree\ , \ 90\degree~~~]$ \\
\hline\hline
\end{tabular}
\label{tab:dataset}
\end{table}

\subsection{Evaluation metric}
Log-spectral distortion (LSD) is employed to evaluate our method:
\begin{equation}
\textit{LSD}(\operatorname{H}, \hat{\operatorname{H}})=\sqrt{\frac{1}{LK} \sum_{\theta, \phi}\sum_{k}\left(20 \log _{10} \left| \frac{\operatorname{H}(\theta, \phi, k)}{\hat{\operatorname{H}}(\theta, \phi, k)} \right| \right)^{2}},
\label{eq: lsd}
\end{equation}
where $k$ denotes the frequency index. $L$ and $K$ are the numbers of spatial locations and frequency bins, respectively. $\operatorname{H}(\theta, \phi, k)$ and $\hat{\operatorname{H}}(\theta, \phi, k)$ indicate the linear-scale magnitude of the ground-truth HRTF and the predicted HRTF of the $k$-th frequency bin, respectively, at the direction of azimuth angle $\theta$ and elevation angle $\phi$.

\subsection{Data preprocessing}
The HRTFs are obtained by taking a 256-point fast Fourier transform (FFT) of the HRIRs. 
To lessen the impact of the HRIR measuring equipment's frequency response limits, following~\cite{grijalva2016manifold, chen2019autoencoding}, we filter the HRTFs to retain frequencies between 200 Hz and 16 kHz. This also aligns with the fact that frequencies up to 16 kHz contribute to sound localization~\cite{blauert1997psychophysics}. This leaves $K = 92$ frequency bins. 

We unify the range of the azimuth angle to [0\degree, 360\degree) by taking the positive remainder divided by 360. Following~\cite{chen2019autoencoding}, we redefine the azimuth angles of all the right ears as $\theta' = 360 - \theta$ so that the left ear and the right ear can share the same model. This essentially doubles the number of subjects as we view all ears as left ears. To ensure the magnitude continuity at the frontal direction, we extend the azimuths to (-30\degree, 390\degree) during training by repeating the HRTF data in the azimuth range of (0\degree, 30\degree) and (330\degree, 360\degree). We then keep only the range of [0\degree, 360\degree) for evaluation.

Since different databases may have different source distances and playback/recording devices, the HRTFs may have different offsets on the dB scale. 
We normalize the magnitude of the database by taking the average energy on the equator (i.e. $\phi = 0$) across all frequencies. The normalized HRTF magnitude can be described as
\begin{equation}
\operatorname{H}(\theta, \phi, k)=\frac{\operatorname{HRTF}(\theta, \phi, k)}{\sqrt{\frac{1}{360 K}\sum_{\theta}\sum_{k} \operatorname{HRTF}(\theta, 0, k)^2 \Delta \theta}},
\label{eq:normalize}
\end{equation}
where $\operatorname{HRTF}(\theta, \phi, k)$ and $\operatorname{H}(\theta, \phi, k)$ uses linear magnitudes. This normalization shifts HRTF magnitudes in the dB scale but maintains variances across subjects and across different frequencies. 


\subsection{Implementation details}
The MLP network we use contains two hidden layers. Each hidden layer contains 2048 nodes. The latent vector $\mathbf{z}$ is a 32-dim vector ($D=32$).
We use Adam optimizer 
and set the batch size as 18. In each batch, we apply padding and masking for loss calculation since HRTF data from various datasets have different lengths of data samples. The learning rate $\mu$ is initially set to $\mu_0=0.0003$ and decay according to $\mu_i = \mu_0 * \frac{1}{1+0.01 i}$ in the $i$-th epoch. Our network is trained for 300 epochs on a single NVIDIA RTX 2080 Ti GPU. 

\section{Results and Discussions}
With the neural fields representation that is agnostic to sampling schemes, we are able to train on any of the datasets or combinations of them. In this section, we demonstrate the result of  HRTF interpolation and generation with our proposed HRTF field.


\begin{figure}[t]
  \centering
  \centerline{\includegraphics[width=0.99\columnwidth]{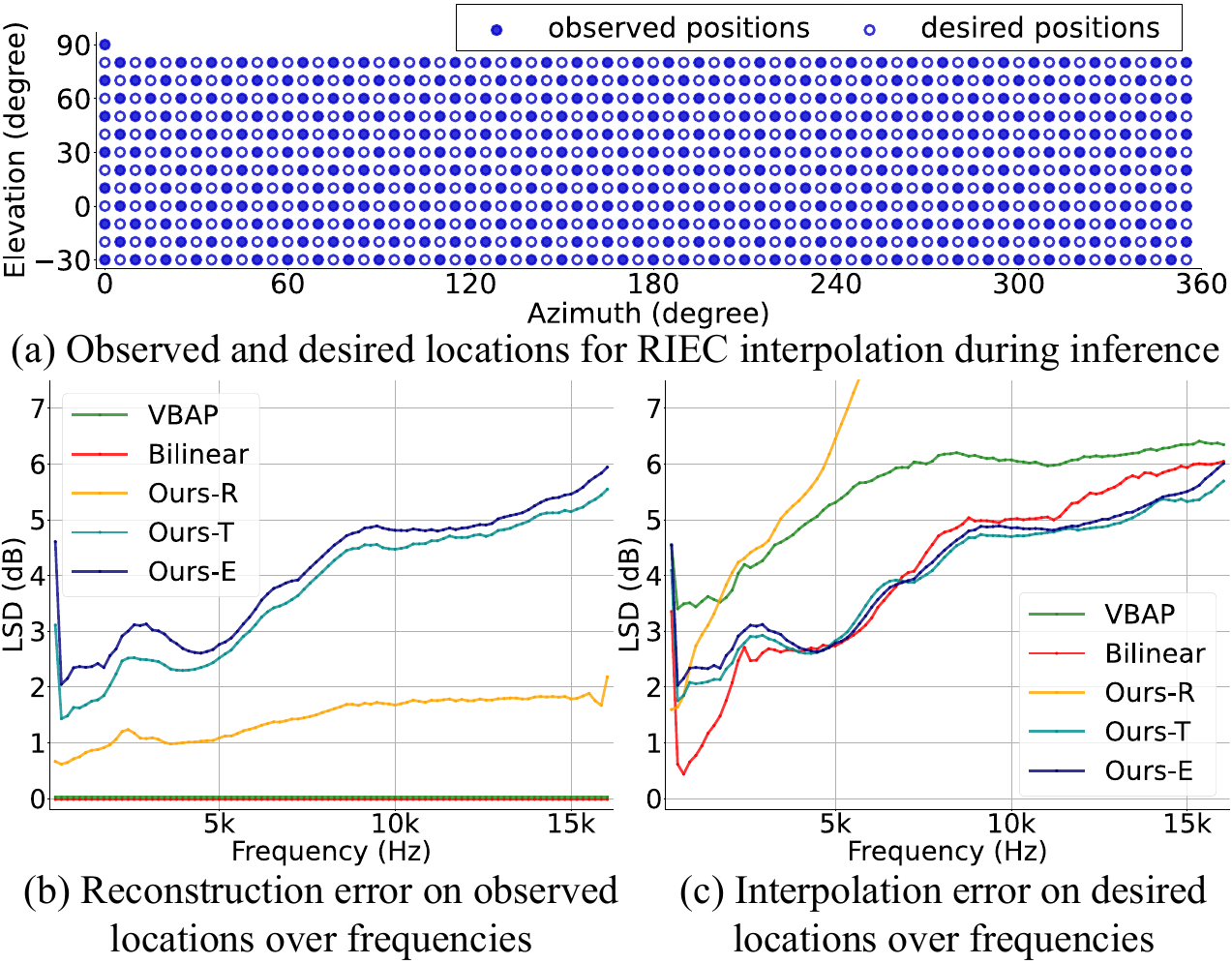}}
\caption{Interpolation setup and the results of reconstruction and interpolation under three of our settings and compare with two baselines.}
\label{fig:interp}
\end{figure}


\subsection{HRTF interpolation with the learned HRTF field}
\label{ssec:interp}
We investigate the location interpolation ability of the proposed method. While our method works on any dataset, we only perform interpolation on the RIEC grids in this section. The observed and desired (i.e., to-be-interpolated) locations are shown in Fig.~\ref{fig:interp} (a). We feed the observed data to different interpolation methods to interpolate the data at the desired locations.

\begin{figure*}[]
  \centering
  \centerline{\includegraphics[width=\textwidth]{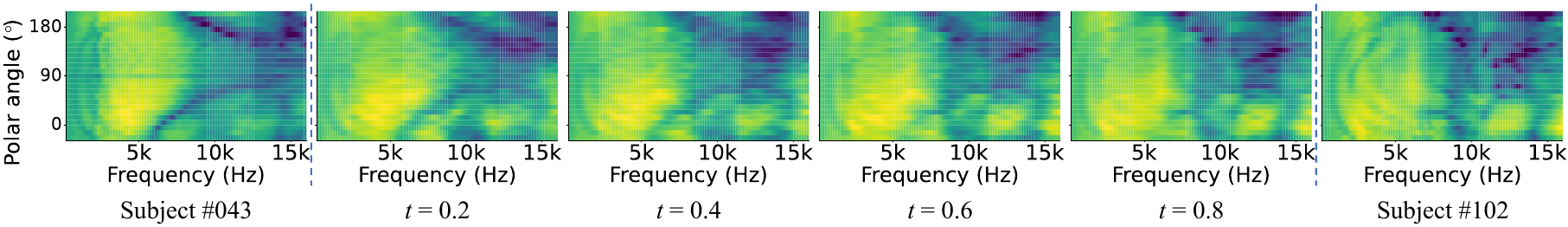}}
\caption{Generation with our proposed HRTF field. All subfigures are HRTF magnitudes (dB) of the midsagittal planes. The leftmost and the rightmost subfigures are normalized HRTFs from two subjects. The middle four subfigures are synthesized from linear interpolation with $t$ as the coefficient between latent codes for the trained IGON.}
\label{fig:gen}
\end{figure*}

\begin{figure}[]
  \centering
  \centerline{\includegraphics[width=0.99\columnwidth]{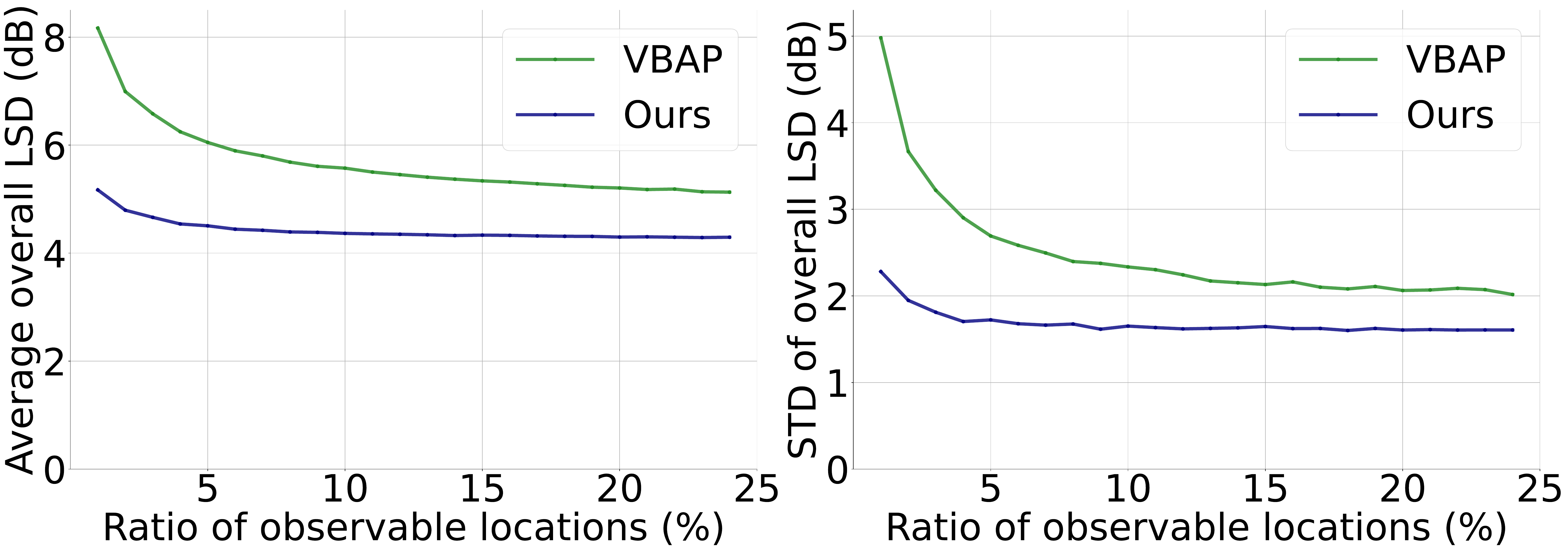}}
\caption{The mean (left) and standard deviation (right) of  overall LSD across all subjects for conditional generation on desired locations.}
\label{fig:cgen}
\end{figure}

We use three settings for our method. The key difference among the three settings is in the training data, that is, whether the test subject (observable locations) or the subject's dataset is used in training. 
(1) \textbf{Ours-R}: The model is trained on the entire RIEC dataset (only using the observable locations). 
(2) \textbf{Ours-T}: The model is trained on the entire RIEC dataset (only using the observable locations) together with all the other nine datasets (using all their locations). 
(3)~\textbf{Ours-E}: The entire RIEC dataset is held out, while only the other nine datasets are used for training. 
For all three settings, the to-be-interpolated positions of the RIEC datasets are held out and only used during evaluation. For evaluation, we compare with the ground-truth magnitude and calculate the error by Eq.~\eqref{eq: lsd}.

For our methods with three settings, we first train an IGON with the training data to learn the HRTF Field and then fix the model. During inference, we calculate $\mathbf{z}$ from the observable locations, then predict the magnitudes for the desired locations.

We adopt two widely used mathematical interpolation methods for comparison, vector base amplitude panning (VBAP)~\cite{pulkki1997virtual} and 3D bilinear interpolation~\cite{freeland2004interpolation}. Both VBAP and bilinear approaches seek multiple neighboring positions and calculate a weighted combination of the neighboring magnitudes. We use the MATLAB function
\href{https://www.mathworks.com/help/audio/ref/interpolatehrtf.html}{\texttt{interpolateHRTF}} 
for implementation of both methods.

Fig.~\ref{fig:interp} (b) and (c) show the reconstruction error and interpolation error, respectively. The bilinear and VBAP have 0 dB error at the source locations, since they only predict the desired locations. 
For our first setting Ours-R, the reconstruction error is low, but the interpolation error is quite high, even higher than 7 dB for frequencies over 5 kHz. This suggests that the model overfits the locations seen during training and fails to predict magnitudes at novel locations. 
With the Ours-T setup, the reconstruction error is higher, showing that combining all the datasets may make the regression difficult. This might suggest that a better normalization or calibration method besides Eq.~\eqref{eq:normalize} is needed to mitigate other factors. For interpolation, it outperforms VBAP and achieves similar results with bilinear between 4-10 kHz, and is better at higher frequencies. This shows that the manifold information is useful for interpolation, especially at high frequencies.
As for the Ours-E setup, the result is similar to Ours-T, with only slight degradation in the reconstruction, showing the cross-dataset generalization ability of our proposed method.


\subsection{Conditional generation from randomly observed locations}
\label{ssec:cond_gen}
Our method maps HRTFs to a latent space. 
If $\z$ can be inferred from geometry of a subject,
we can then use the proposed method to generate the subject's HRTF. As inference of $\z$ from other data is outside the scope of this paper, here we validate the HRTF generation ability from $\z$ that is inferred from a partial observation of HRTF data at a few locations. 
This is similar to location interpolation in Section~\ref{ssec:interp}, but the observed locations are much fewer and irregular.

We perform a conditional generation on the RIEC dataset with the IGON model trained on the other nine datasets (the same as the Ours-E setup in Section~\ref{ssec:interp}). We randomly sample some percentage (less than 25\%) of the locations and hold the rest of the locations for prediction.
We use VBAP and bilinear interpolation methods for comparison.

The average and standard deviation of overall LSD result (calculated across all frequencies on desired locations) across all subjects in the RIEC dataset is shown in Fig.~\ref{fig:cgen}. The bilinear method collapses since the observable locations are too sparse. Both VBAP and our method work, but ours is superior to VBAP in both average and standard deviation. This demonstrates the advantage of the proposed learning-based method that can leverage information from other subjects (even from other datasets) on HRTF generation. This advantage would be better demonstrated when $\z$ is inferred from the head geometry of the test subject, as mentioned earlier. We believe this could be promising future work.

\subsection{Generation by sampling from the latent space}
We further demonstrate the generation ability of the proposed HRTF field on an unconditional generation task. We train our model on the entire RIEC dataset to learn the latent $\mathbf{z}$ space over the HRTFs. We sample two latent codes $\z_1, \z_2$ of two subjects' HRTFs. 
The latent codes are computed according to Eq.~\eqref{eq:z0}. Then we linearly combine the two latent codes by $\z_t = (1 - t) \z_1 + t \z_2$, where $t$ is the weight parameter. We synthesize new HRTF samples with HRTF field $G(~\boldsymbol{\cdot}, ~\boldsymbol{\cdot}, \z_t)$, and their midsagittal planes are shown in Fig.~\ref{fig:gen}.


The overall patterns of the subfigures are similar, showing that the generated HRTF samples generally follow the HRTF manifold. When comparing the leftmost two or the rightmost two, where a slight change is made in the latent space, the spectral cues (notches) are well maintained in the median plane, suggesting that the generated HRTFs well preserve the spectral cues. The transition of the middle synthesized four subfigures from left to right is smooth.

This could be useful for selection-based HRTF personalization. Instead of choosing from measured HRTFs of existing subjects  with the best matching geometry~\cite{cobos2022overview}, we can interpolate the latent codes among existing subjects to synthesize new personalized HRTFs.










\section{Conclusions}
\label{sec:cls}
We propose an implicit, unified, and differentiable representation for HRTF, learned by a deep neural network from discrete samples of a function defined in a continuous space. We further propose HRTF field,  a generative model to generate neural fields representations from the latent space. The generative model performs well in interpolation and generation tasks.
We believe that such a generative method will significantly advance data-driven research on HRTF personalization.
Future work includes HRTF phase or direct HRIR modeling with neural fields, and integrating HRTF field with physical laws to leverage the differentiable property of neural fields.

\vfill\pagebreak

\footnotesize


\bibliographystyle{IEEEbib}
\bibliography{refs}

\end{document}